\documentclass[journal=jacsat,manuscript=article]{achemso}

\usepackage{chemformula} 
\usepackage{hyperref}
\usepackage[T1]{fontenc} 
\usepackage{subcaption}
\usepackage{titlesec}

\author{Prabeen Kumar Pattnayak$^1$}
\author{Aloke Kumar$^1$}
\author{Gaurav Tomar$^1$}
\email{*gtom@iisc.ac.in}
\affiliation[Indian Institute of Science]
{$^1$Department of Mechanical Engineering, Indian Institute of Science, CV Raman road, Bengaluru, Karnataka-560012, India}

\title[Re-orientational dynamics of ring polymers in dilute solutions]
  {Re-orientational dynamics of ring polymers in dilute solutions}

\abbreviations{IR,NMR,UV}

\begin{document}



\begin{abstract}

Advances in controlled polymerization have enabled the synthesis of mechanically interlocked polymers like molecular knots and linear[n]catenane. These aesthetic macromolecules with unique topological constraints in the form of mechanical bonds are well known for their fascinating transport and rheological properties in the development of molecular machines and in knotted protein dynamics in biological applications. The diffusion dynamics of such macromolecular structures with large internal degrees of freedom are generally studied by using an equivalent size parameter, i.e., hydrodynamic radius, defined using Zimm theory. 
Although diffusion rates are expected to depend strongly on the molecular topological constraints in macromolecules, their explicit effects on translational and reorientational dynamics are still unknown. Here, we perform an \textit{in silico} study on the diffusion dynamics of seven topologically distinct polymer chains in the limit of infinite dilution using multi-particle collision dynamics. The modeled polymers are linear, ring, linear[2]catenane, trefoil knot, linear[3]catenane, cyclic[3]catenane, and Borromean ring. The molecular weights of these macromolecules are selected such that the resulting hydrodynamic radius is approximately equal to each other. We show that while the translational diffusion coefficients of these topologically distinct polymer chains are approximately equal to each other in agreement with the Zimm theory, there are significant differences among the values of the corresponding rotational diffusion coefficients. We show that the presence of mechanical bonds in the polymer chains slows down the rotational diffusion significantly, thus suggesting the role of molecular topology on reaction kinetics of macromolecules. 

\end{abstract}


\section*{Significance statement}
Polymer chains constituting mechanically interlocked rings, such as, linear[n]catenanes, and having knotted structures like trefoil knot generally referred to as topologically complex macromolecules. We investigate the rotational and translational diffusion properties of seven topologically distinct macromolecules undergoing Brownian motion in the dilute limit. We show that the mechanical bonds present in the macromolecules tend to slow down their reorientation rate. The results from the present work would be of potential practical value in various fields where rotational diffusion of macromolecules is significant, such as in the fingerprinting of single proteins using a nanopore, in the kinetics of protein-protein self-assembly, in macromolecular crowding, and in the design of molecular machines.

\section*{Introduction}
Brownian motion serves as a connecting link between the physical and biological sciences, observed in phenomena ranging from the random jittery movement of simple pollen grains in water\cite{brown1828xxvii} to the intricate thermal motion of complex macromolecules within living systems. The diffusion dynamics of long chain molecules is one of the fundamental problems in soft matter physics which has broad scientific importance, for example, in DNA dynamics\cite{cohen2007principal,gong2014translational,perkins1994direct,bravo2017supramolecular,smith1995self,kundukad2010control}, protein folding\cite{neuweiler2009direct,dobson2003protein,matthews1993pathways}, polymer nanocomposites\cite{rostom2020polymer,gam2011macromolecular}, polymer-dispersed liquid crystals\cite{bunning2000holographic,coates1995polymer}, and drug delivery\cite{wolak2013diffusion}. Over the past two decades, there has been remarkable progress in synthesizing mechanically interlocked polymers like linear[n]catenane, trefoil knot, cyclic[3]catenane, and many more\cite{sauvage2008molecular,sauvage2017chemical}. These unique macromolecules are well known for their aesthetic topological structures as well as their significance in a plethora of engineering applications, which include drug delivery\cite{coti2009mechanised}, molecular machines\cite{sauvage2008molecular}, catalysis\cite{thordarson2003epoxidation}, and switchable surfaces\cite{davis2010mechanically}. Polymer chains with distinct molecular topological constraints have been reported to have peculiar transport and rheological properties like power law viscoelastic stress relaxation of polymer molecules with ring topology\cite{kapnistos2008unexpected}, non-monotonic dependence of viscosity on ring size in poly[n]catenane melts\cite{rauscher2020dynamics}, and a 100-fold decrease in the translational diffusion coefficients of DNA ring molecules by changing the topology of the surrounding DNA molecules from ring to linear\cite{robertson2007strong}.

The diffusion dynamics of such macromolecular structures with large internal degrees of freedom is generally studied using an equivalent size parameter, i.e., hydrodynamic radius ($R_h$). It is defined by preaveraging the pairwise hydrodynamic interactions in the Zimm theory\cite{dunweg1991microscopic,dunweg1993molecular,dunweg1998molecular}. The relation between the translational diffusion coefficient ($D$) of a polymer chain with its hydrodynamic radius is given by Stokes-Einstein relation\cite{einstein1905molekularkinetischen,wei2020coarse}, $D \sim k_B T /6 \pi \eta R_h$, and that of the rotational diffusion coefficient ($D_r$) is given by Stokes-Einstein-Debye relation\cite{debye1929polar,wei2020coarse}, $D_r \sim k_B T / 8 \pi \eta R_h^3$, where $k_B$ is the Boltzmann constant, $T$ is temperature, and $\eta$ is the dynamic viscosity of the solvent.

One of the intriguing questions in soft matter physics is the effect of the topological constraints on the diffusion dynamics of a polymer chain. There have been some seminal works in this regard. 
Using fluorescence microscopy, Robertson et al.\cite{robertson2006diffusion} calculated the translational diffusion coefficients of DNA molecules with three distinct topological constraints, which are linear, relaxed circular, and supercoiled circular of the same molecular weight, and have reported the supercoiled DNA having the highest diffusion coefficient, while linear DNA has the lowest corresponding value. Kanaeda and Deguchi\cite{kanaeda2009universality} have reported higher values of translational diffusion coefficients for trefoil knot macromolecules than ring and linear ones of the same molecular weights using Brownian dynamics simulations with hydrodynamic interactions. However, in such comparative studies of the topologically distinct macromolecules having the same molecular weight, the explicit effect of the topological constraints (if there are any) on the diffusion dynamics of the polymer chains is dominated by the resulting size differences (in terms of hydrodynamic radius) among the macromolecules. Thus, it is essential to do a comparative study on the diffusion dynamics of polymers with distinct topological constraints at the same hydrodynamic radius.

\begin{figure}[t!]
\centering
\includegraphics[width=0.95\linewidth]{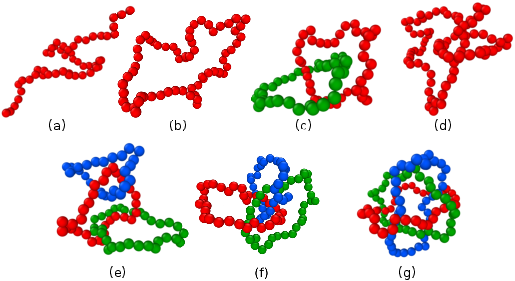}
\caption{Schematic representation of topologically distinct polymer chains: (a) linear chain, (b) ring, (c) linear[2]catenane, (d) trefoil knot, (e) linear[3]catenane, (f) cyclic[3]catenane, and (g) Borromean ring. All seven polymer chains have approximately the same hydrodynamic radius ($R_h \sim 10 \sigma_p$).}
\label{fig:schematic_topology}
\end{figure}

The objective of the present work is to study the essential features of the diffusion dynamics, both translational and rotational, of macromolecules with distinct topological constraints. In spite of a growing interest in topologically complex polymers\cite{farimani2024effects,chen2021entropic,bonato2022topological,li2024self,li2024activity,chiarantoni2025effect,mei2024unified}, diffusion coefficients are still investigated using hydrodynamic radius as the sole parameter. Here, we delineate the effect of various types of mechanical bonds by studying the dynamics of polymeric molecules with different topologies in a dilute solution but with the same hydrodynamic radius. The present study would be of broader practical interest considering the fact that the rotational diffusion is a critical parameter in the fingerprinting of single proteins using a nanopore\cite{yusko2017real}, in the kinetics of protein-protein self assembly\cite{northrup1992kinetics,newton2015rotational}, in studying the local viscosity of a liquid along with its inhomogeneity\cite{zondervan2007local}, and in macromolecular crowding\cite{li2009translational}. Molecular dynamics (MD) is an ideal tool for obtaining insights into the dynamic behavior of polymer chains with complex molecular topology undergoing Brownian motion. We perform an \textit{in silico} study on the diffusion dynamics of seven topologically distinct macromolecules (with approximately the same hydrodynamic radius) in the limit of infinite dilution (see schematic in Figure \ref{fig:schematic_topology}). These polymers are linear, ring, linear[2]catenane, trefoil knot, linear[3]catenane, cyclic[3]catenane, and Borromean ring. The solvent particles are modeled using a mesoscopic coarse-grained simulation method, namely multi-particle collision dynamics (MPCD)\cite{malevanets1999mesoscopic,allahyarov2002mesoscopic}, as it consolidates both thermal fluctuations and long-range hydrodynamic interactions\cite{gompper2009multi}. The measured parameters are expressed using the reduced units: energy scale $k_B T$, length scale $\sigma_p$ ($\sigma_p$ is the diameter of a single monomer), and time scale $\tau$ ($\tau = \sqrt{m \sigma_p^2 / k_B T}$, $m$ is the mass of a solvent particle). 

\section*{Results and Discussion}
Comparative studies on translational diffusion of linear and ring polymer at approximately constant $R_g$ by Hegde et al.\cite{hegde2011conformation} using three different simulation techniques result in a higher translational diffusion coefficient of linear polymer than the ring one, which confirms that $R_h$ is a more appropriate size parameter than $R_g$ in the diffusion dynamics of macromolecules. The variation in $R_g$ with $N$ for the polymer chains under consideration is discussed in section I of the supporting information. The hydrodynamic radius, $R_h$, of the polymer chains is calculated as follows\cite{dunweg1991microscopic,dunweg1993molecular,dunweg1998molecular},
\begin{equation} \label{eqn:rh}
   \langle \frac{1}{R_h} \rangle = \frac{1}{N^2} \sum_{i \ne j} \langle \frac{1}{r_{ij}} \rangle
\end{equation}
where $r_{ij}$ is the distance between the monomers and $N$ is the molecular weight of the corresponding polymer chain. The molecular weights of the seven topologically distinct polymer chains are selected such that the resulting hydrodynamic radius is approximately the same ($R_h \sim 10 \sigma_p$), as shown in Table \ref{tbl:10rh}.

We perform hybrid MD-MPCD calculations using the excluded volume interactions between monomers, and the bonds between the monomers are modeled using FENE springs and an angular potential (see Materials and Methods section for details). We study the variations in the center-of-mass and polymer chain orientation with time to determine the translational and rotational diffusion coefficients.

\subsection{Translational diffusion}
\begin{figure}
\centering
\includegraphics[width=0.95\linewidth]{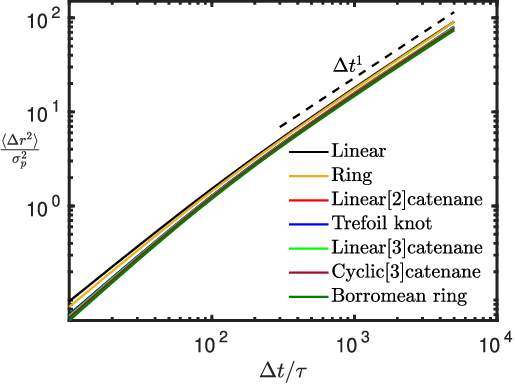}
\caption{Variation in the center-of-mass mean square displacement ($\Delta r^2$) with lag-time ($\Delta t$) for seven topologically distinct polymer chains having approximately the same hydrodynamic radius ($R_h \sim 10 \sigma_p$).}
\label{fig:msd}
\end{figure}

The center-of-mass mean square displacement ($\Delta r^2$) vs. lag-time ($\Delta t$) plots for all seven polymers are shown in Figure \ref{fig:msd}. After the inertial regime ($\Delta t < 300 \tau$), the linear diffusive regime is reached, from which the translational diffusion coefficients of the respective polymer chains are calculated by using the relation, $\Delta r^2 = 6 D \Delta t$, and are summarized in the second last column of Table \ref{tbl:10rh}. The differences in the values of the translational diffusion coefficients among the seven topologically distinct polymer chains having approximately the same value of $R_h$ are found to be almost negligible. This is in good agreement with the Zimm theory\cite{dunweg1998molecular}, which states that the translational diffusion coefficient of a polymer chain is inversely proportional to its hydrodynamic radius. Thus, simulation of polymer chains using MPCD yields Zimm-like dynamics, which has also been demonstrated earlier in literature using the linear chain with and without excluded volume\cite{mussawisade2005dynamics} and using the fully flexible star-shaped polymers\cite{pattnayak2024diffusion}. Direct comparison with Zimm theory using the Kirkwood formula is discussed in section B of the supporting information.
These results indicate that molecular topological constraints do not have any additional influence on the translational diffusion of the polymer chains, besides determining their hydrodynamic radius. Thus, the hydrodynamic radius is a sufficient parameter for studying translational diffusion dynamics of polymer chains with complex topological constraints.

\subsection{Rotational diffusion}
\begin{figure}[b!]
\centering
\includegraphics[width=0.8\linewidth]{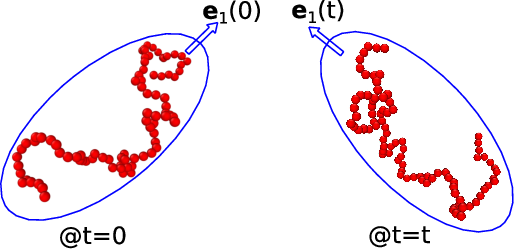}
\caption{Schematic illustration of re-orientation of a polymer chain using an imaginary ellipsoid. The normalized eigenvector ($\boldsymbol{e_1}$), corresponding to the largest eigenvalue of the gyration tensor, is used to measure the reorientation rate of the polymer chain.}
\label{fig:ellip}
\end{figure}

\begin{figure}
\centering
\includegraphics[width=0.95\linewidth]{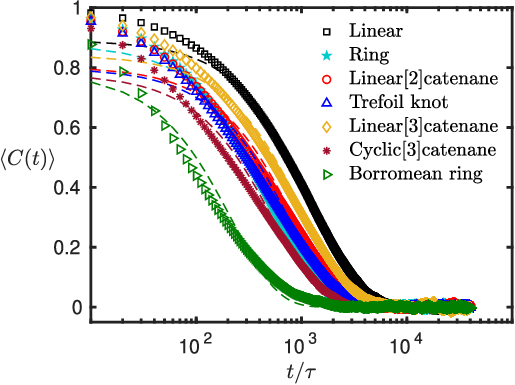}
\caption{Variation in reorientation correlation function ($C(t)$) with time for the seven topologically distinct polymer chains having approximately the same hydrodynamic radius ($R_h \sim 10 \sigma_p$). The dashed lines represent single exponential fits using the least squares method. The corresponding coefficients of determination are 0.99 (linear), 0.99 (ring), 0.99 (linear[2]catenane), 0.99 (trefoil knot), 0.99 (linear[3]catenane), 0.98 (cyclic[3]catenane), and 0.97 (Borromean ring).}
\label{fig:rcf}
\end{figure}

Polymer chains reorient themselves continuously while simultaneously diffusing translationally. The reorientation of a polymer chain can be considered analogous to the rotation of an equivalent imaginary ellipsoid surrounding the polymer chain, as shown schematically in Figure \ref{fig:ellip}. The magnitudes and directions of the semi-axes of this imaginary ellipsoid are given by the three eigenvalues and the three mutually orthogonal eigenvectors of the gyration tensor\cite{theodorou1985shape} of the polymer chain, respectively. The rate of reorientation can be measured by any vector rigidly attached to the polymer chain. Here, we use the normalized eigenvector, $\boldsymbol{e_1}$, which corresponds to the largest eigenvalue of the gyration tensor of the polymer chain. The corresponding reorientational correlation function (RCF)\cite{wong2009influence} is defined as,
\begin{equation}
   C(t) = \langle P_2( \boldsymbol{e_1}(0).\boldsymbol{e_1}(t) ) \rangle
\end{equation}
where $P_2(x) = (3x^2 - 1)/2$ is the second-order Legendre polynomial, and the angle bracket represents the time and ensemble average over five system replicas. The decays of the respective RCFs of the seven topologically distinct polymer chains with time are shown in Figure \ref{fig:rcf}. The faster the decay of the RCF, the higher the rate of reorientation of the corresponding polymer chain. The RCF of the Borromean ring decays the fastest, while that of the linear chain decays the slowest. The rotational diffusion coefficient can be calculated by fitting a single exponential to the decay of RCF\cite{wong2009influence}, i.e., $C(t) = e^{-6 D_r t}$. The computed values of $D_r$ for the seven polymers under consideration are shown in the last column of Table \ref{tbl:10rh}. Comparison of the computed values of $D_r$ from the present simulation with Perrin’s formula\cite{perrin1934movement,perrin1936movement,ryabov2006efficient} is shown in section E of the supporting information.

\begin{table*}\centering
\caption{Molecular weight ($N$), Radius of gyration ($\bar{R_g} = \langle R_g^2 \rangle^{1/2}$), Hydrodynamic radius ($\bar{R_h} = \langle 1 / R_h \rangle^{-1} $),  Relative shape anisotropy($\kappa^2$), Translational diffusion coefficient ($D$), and Rotational diffusion coefficient ($D_r$) of polymer chains with distinct topology having approximately the same values of the hydrodynamic radius ($R_h \sim 10 \sigma_p$).}

\begin{tabular}{lcccccc}
Type & $N$ & $\bar{R_g}/ \sigma_p$ & $\bar{R_h} / \sigma_p$ & $\kappa^2$ & $D \tau / \sigma_p^2$ & $D_r \tau $ \\
1. Linear & $52$ & $5.52 \pm 1.05$ & $10.06 \pm 1.86$ &  $0.46 \pm 0.194$ & $3 \times 10^{-3}$ & $1.23 \times 10^{-4}$ \\
2. Ring & $64$ & $4.69 \pm 0.55$ & $9.97 \pm 2.02$ & $0.274 \pm 0.124$ & $3 \times 10^{-3}$ & $2.86 \times 10^{-4}$ \\
3. Linear[2]catenane & $84$ & $4.6 \pm 0.43$ & $10.09 \pm 1.95$ & $0.201 \pm 0.117$ & $2.7 \times 10^{-3}$ & $1.96 \times 10^{-4}$ \\
4. Trefoil knot & $88$ & $4.49 \pm 0.42$ & $10 \pm 2.06$ & $0.196 \pm 0.105$ & $2.7 \times 10^{-3}$ & $2.19 \times 10^{-4}$ \\
5. Linear[3]catenane & $93$ & $4.44 \pm 0.39$ & $10.06 \pm 2.2$ & $0.218 \pm 0.119$ & $2.7 \times 10^{-3}$ & $1.58 \times 10^{-4}$ \\
6. Cyclic[3]catenane & $102$ & $4.22 \pm 0.25$ & $9.95 \pm 2.13$ & $0.115 \pm 0.068$ & $2.6 \times 10^{-3}$ & $3.04 \times 10^{-4}$ \\
7. Borromean ring & $114$ & $4.18 \pm 0.21$ & $10.05 \pm 2.11$ & $0.07 \pm 0.048$ & $2.5 \times 10^{-3}$ & $7.6 \times 10^{-4}$ \\
\end{tabular}
\label{tbl:10rh}
\end{table*}

Unlike for translational diffusion, the computed values of the rotational diffusion coefficients of most of the polymer chains are significantly different from each other, even though their hydrodynamic radius values are maintained approximately constant. Since the equivalent size parameter of the seven polymer chains is fixed, the differences in the values of $D_r$ can possibly be analyzed by considering the other two geometrical parameters of the polymer chains, i.e., shape and topology. The eigenvalues of the gyration tensor are useful in defining multiple shape parameters\cite{vsolc1971shape,theodorou1985shape,khabaz2014effect}, out of which relative shape anisotropy can be expressed in terms of the invariants of the gyration tensor of the polymer chain and is the overall measure of the shape anisotropy\cite{theodorou1985shape,khabaz2014effect}. The relative shape anisotropy ($\kappa^2$) is calculated as follows,
\begin{equation} \label{eq:shape}
    \kappa^2 = 1 - 3 \frac{\lambda_1 \lambda_2 + \lambda_2 \lambda_3 + \lambda_3 \lambda_1}{( \lambda_1 + \lambda_2 + \lambda_3 )^2}
\end{equation}
where $\lambda_1$, $\lambda_2$, and $\lambda_3$ are the eigenvalues of the gyration tensor of the polymer chain. $\kappa^2$ ranges from 0 (represents a sphere or any of the five Platonic solids) to 1 (represents a straight rod).
The relative shape anisotropy values of the seven topologically distinct polymer chains are calculated from the simulation and summarized in the fifth column of Table \ref{tbl:10rh}. The linear chain has the highest value of relative shape anisotropy, whilst the Borromean ring has the lowest value. The variation of $\kappa^2$ with $N$ for the polymer chains under consideration is discussed in section K of the supporting information.

\begin{figure}[b!]
\centering
\includegraphics[width=0.95\linewidth]{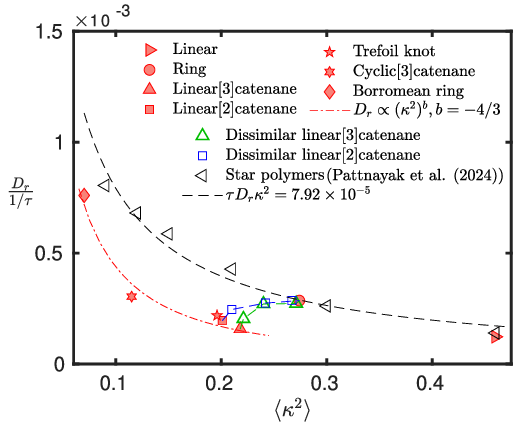}
\caption{Rotational diffusion coefficient ($D_r$) vs. relative shape anisotropy ($\kappa^2$) at approximately constant hydrodynamic radius: $R_h \sim 10 \sigma_p$(Present work). The seven topologically distinct polymers (red markers) are represented by their respective values of $\kappa^2$: 0.46 (linear), 0.274 (ring), 0.218 (linear[3]catenane), 0.201 (linear[2]catenane), 0.196 (trefoil knot), 0.115 (cyclic[3]catenane), and 0.07 (Borromean ring). The red dotted line represents a power-law fit using the least squares method with a coefficient of determination 0.97. The dissimilar linear[3]catenanes (green upward triangles) are represented by their respective values of $\kappa^2$: 0.221 (44,24,24), 0.24 (50,18,18), and 0.271 (58,10,10). The dissimilar linear[2]catenanes (blue squares) are represented by their respective values of $\kappa^2$: 0.21 (48,32), 0.242 (56,22), and 0.267 (60,12). The results of the star polymers (Pattnayak et al.\cite{pattnayak2024diffusion}) at constant hydrodynamic radius: $R_h \sim 9.56 \sigma_p$ are represented by black left-pointing triangles, where the star polymers are represented by their respective values of $\kappa^2$: 0.46 (linear), 0.3 (3 armed star), 0.21 (4 armed star), 0.15 (5 armed star), 0.12 (6 armed star), and 0.09 (7 armed star). The black dashed line represents the rectangular hyperbola, which was observed for the case of star polymers (Pattnayak et al.\cite{pattnayak2024diffusion}).}
\label{fig:hyperbola}
\end{figure}

The rotational diffusion coefficient vs. relative shape anisotropy plot for the seven polymers under consideration with approximately constant $R_h$ is shown in Figure \ref{fig:hyperbola}. The polymers are represented by their respective values of $\kappa^2$. The results for the star polymers  are shown for reference here \cite{pattnayak2024diffusion}. The star polymers with various numbers of arms and having approximately the same hydrodynamic radius ($R_h \sim 9.56 \sigma_p$) are represented by their respective $\kappa^2$ values. The star polymers' results suggest that a lower value of relative shape anisotropy leads to a higher rotational diffusion coefficient of the polymer chain at constant hydrodynamic radius\cite{pattnayak2024diffusion}. This inverse relation between $D_r$ and $\kappa^2$, i.e., $D_r \sim 1/\kappa^2$ (at constant $R_h$), is approximated by a rectangular hyperbola, $\kappa^2 D_r = 7.92 \times 10^{-5}$ in Pattnayak et al.\cite{pattnayak2024diffusion}, and is shown using the black dashed line in Figure \ref{fig:hyperbola}. The results of the present work (red markers) indicate that the relation between $D_r$ and $\kappa^2$ (at constant $R_h$) for topologically complex polymer chains is not as straightforward as in the case of star polymers. 
Additional constraints in the case of more complex topologies considered here, in contrast to star polymers, results in significantly lower rotational diffusivity for the same shape anisotropy.

In what follows, we examine the effect of MPCD solvent properties 
and the flexibility of the polymer chain on $D_r$.
 We note that the small difference in $R_h$ $\sim 0.44 \sigma_p$ (where $R_h \sim  9.56\sigma_p - 10\sigma_p$) between the polymers under consideration and the star polymers \cite{pattnayak2024diffusion}, the values of $D_r$ of the linear chain in both the cases are very close to each other, as can be seen in Figure \ref{fig:hyperbola}. Thus, the small size difference of $0.44 \sigma_p$ has very minimal effect on the reorientation dynamics of the polymer chains, which can be neglected. The transport properties of the solvent have a strong influence on the values of the diffusion coefficients ($D,~D_r$) of the polymer chains. MPCD fluid transport properties are functions of collision cell length, collision time step, and solvent particle density\cite{allahyarov2002mesoscopic}. In the present work, the values of these parameters are selected to be exactly the same as we have considered in \cite{pattnayak2024diffusion}, which eventually results in the same value of dynamic viscosity and Schmidt number of the MPCD fluids in both cases.

   
     Finally, the chain flexibility of equally-sized polymers can also affect the reorientation dynamics. In the present work, we have given the standard angular potential (see equation \ref{eqn:angle}) among the adjacent monomers for all the polymer chains, making these semi-flexible. So, in order to compare the effect of the flexibility, we perform simulations of these seven topologically distinct polymers by removing the angular potential, thus making these fully flexible, and maintaining approximately the same size ($R_h \sim 10 \sigma_p$). The plot of RCF decay with time of all seven polymers for both fully flexible and semi-flexible cases is given in Figure S4 in the supporting information. These plots show that the RCF curves of both the semi-flexible and the fully flexible polymers coincide very well with each other for linear, ring, linear[2]catenane, and trefoil knot, and are close to each other for linear[3]catenane, cyclic[3]catenane, and Borromean ring. Thus, the effect of flexibility on the reorientation dynamics of polymers with approximately the same values of the hydrodynamic radius is negligible. 

Hence, we conclude that any reasonable deviation in values of $D_r$ from the rectangular hyperbola (see Figure \ref{fig:hyperbola}) can only be attributed to the presence of the topological constraints in the polymer chains. Interestingly, the three catenanes, trefoil knot, and Borromean ring, all lie on the curve $<\kappa^2>^{-4/3}$, indicating commonality in the nature of constraint between these multi-ring type polymer chains.

The $D_r$ values of linear and ring follow the rectangular hyperbola very well, as shown in Figure \ref{fig:hyperbola}. The linear chain, being free from any topological constraints, has its $D_r$ value lying very close to the rectangular hyperbola. The ring result suggests that, even though there is a simple topological constraint of one closed end in the ring polymer, it does not explicitly affect the reorientation dynamics and its effect is implicitly taken care of by the resulting value of $\kappa^2$. In the case of the other five configurations, i.e., linear[3]catenane, linear[2]catenane, trefoil knot, cyclic[3]catenane, and Borromean ring, there are various types of mechanical bonds present in the structures. Thus, these are topologically more constrained than the ring polymer. The corresponding values of $D_r$ for these five cases are significantly less than the corresponding values as per predictions of the rectangular hyperbola using their respective $\kappa^2$ values, as can be seen in Figure \ref{fig:hyperbola}. For example, the values of $D_r$ of linear[3]catenane ($1.58 \times 10^{-4} \tau^{-1}$), linear[2]catenane ($1.96 \times 10^{-4} \tau^{-1}$), and trefoil knot ($2.19 \times 10^{-4} \tau^{-1}$) are reasonably less than that of the 4-arm star polymer ($4.28 \times 10^{-4} \tau^{-1}$), even though their $\kappa^2$ values, 0.218 (linear[3]catenane), 0.201 (linear[2]catenane), and 0.196 (trefoil knot) are very close to that of the 4-arm star polymer ($0.21$). The $D_r$ value of cyclic[3]catenane is $3.04 \times 10^{-4} \tau^{-1}$, which is significantly lower than that of 6-arm star polymer ($D_r = 6.81 \times 10^{-4} \tau^{-1}$), despite having a value of $\kappa^2 = 0.115$ being very close to that of the 6-arm star polymer ($\kappa^2 = 0.12$). These results indicate that the rotational diffusion coefficient of a polymer chain having complex topological constraints in the form of mechanical bonds is lower than that of a polymer chain without any topological constraints of approximately equal shape ($\kappa^2$) and size ($R_h$). Thus, we can argue that the presence of mechanical bonds in the polymer chain slows down its rate of reorientation. To confirm this, we consider the following two comparative cases of dissimilar linear[2]catenane and linear[3]catenane, where we regulate the effect of the topological constraints (mechanical bonds) on the reorientation dynamics by adjusting the molecular weights of the corresponding constituent rings.

\begin{figure*}
\centering
\includegraphics[width=0.95\linewidth]{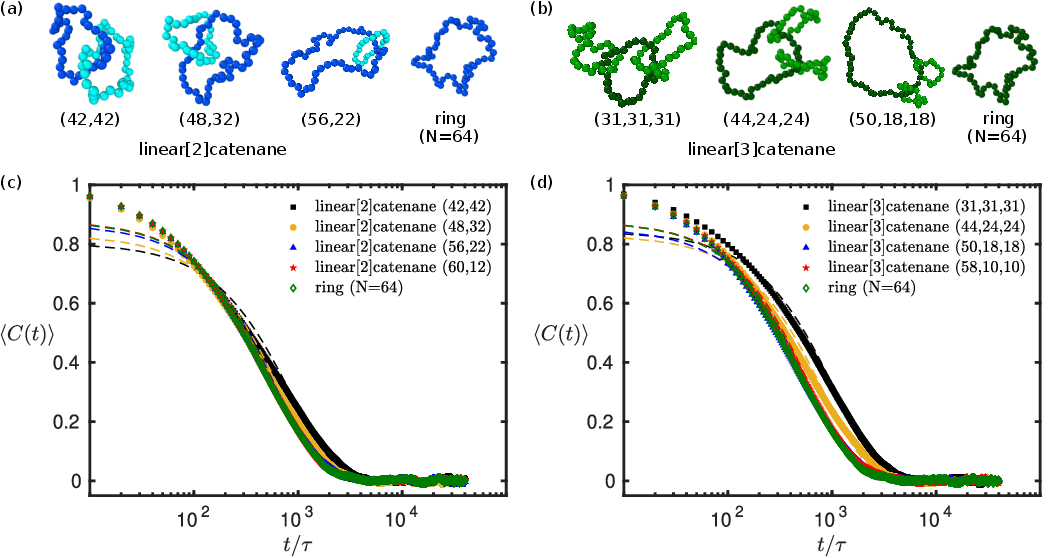}
\caption{(a) Schematic representation of linear[2]catenanes and ring polymer having approximately the same value of hydrodynamic radius ($R_h \sim 10 \sigma_p$). (b) Schematic representation of linear[3]catenanes and ring polymer having approximately the same value of hydrodynamic radius ($R_h \sim 10 \sigma_p$). (c) RCF decay with time for linear[2]catenanes and ring polymer having approximately the same value of hydrodynamic radius ($R_h \sim 10 \sigma_p$).(d) RCF decay with time for linear[3]catenanes with hydrodynamic radius, $R_h \sim 10 \sigma_p$.}
\label{fig:dissimilar}
\end{figure*}

\subsubsection{Case 1: Dissimilar linear[2]catenane} \label{sec:l2c}
Here, we consider the reorientation dynamics of three linear[2]catenane polymers with dissimilar constituent rings, i.e., the molecular weights of the individual constituent rings are different from each other. These polymers are identified with the molecular weights of their respective constituent rings: (48,32), (56,22), and (60,12), as shown in Figure \ref{fig:dissimilar}(a). The molecular weights of the individual constituent rings have been chosen such that the resulting hydrodynamic radius (calculated using equation(\ref{eqn:rh})) of the entire linear[2]catenane is approximately equal to $10 \sigma_p$. The decay of the RCF of these three dissimilar linear[2]catenane with time are shown in Figure \ref{fig:dissimilar}(c) along with that of linear[2]catenane having identical constituent rings, i.e., (42,42) and the ring polymer (N=64), both considered earlier in this work. We note that as the relative size of one of the component rings in linear[2]catenane is decreased, keeping the $R_h$ same, the RCF shifts towards that for the ring polymer (marked in green in the figure) and nearly coincides for the cases of (56,22) and (60,12). Thus, the reorientation dynamics of these two dissimilar linear[2]catenanes is essentially dominated by the respective larger constituent ring due to the reduction in the conformational constraint. As a consequence, the behavior becomes ring-like. This gradual transition in the reorientation dynamics of the linear[2]catenane into ring-like behavior can also be observed from the corresponding values of $D_r$, as shown in the last column of Table \ref{tbl:l2c}, and is represented by the blue dashed line in Figure \ref{fig:hyperbola}, where the dissimilar linear[2]catenane polymers are represented by their respective values of $\kappa^2$.

\begin{table}
\centering
\caption{Molecular weights of the constituent rings $\boldsymbol{(N_1,N_2)}$, Hydrodynamic radius ($\boldsymbol{\bar{R_h} = \langle 1 / R_h \rangle^{-1}}$), Relative shape anisotropy ($\boldsymbol{\kappa^2}$), and Rotational diffusion coefficient ($\boldsymbol{D_r}$) for linear[2]catenanes having approximately the same hydrodynamic radius ($\boldsymbol{R_h \sim 10 \sigma_p}$).}

\begin{tabular}{cccc}
$\boldsymbol{(N_1,N_2)}$ & $\boldsymbol{\bar{R_h} / \sigma_p}$ & $\boldsymbol{\kappa^2}$ & $\boldsymbol{D_r \tau }$ \\
(42,42) & $10.09 \pm 1.95$ & $0.201 \pm 117$ & $1.96 \times 10^{-4}$ \\
(48,32) & $10.09 \pm 2.12$ & $0.21 \pm 0.112$ & $2.46 \times 10^{-4}$ \\
(56,22) & $10.09 \pm 2.02$& $0.242 \pm 0.111$ & $2.75 \times 10^{-4}$ \\
(60,12) & $9.93 \pm 1.98$ & $0.267 \pm 0.118$ &  $2.86 \times 10^{-4}$ \\
Ring ($N=64$) & $9.97 \pm 2.02$ & $0.274 \pm 0.124$ & $2.86 \times 10^{-4}$ \\
\end{tabular}
\label{tbl:l2c}
\end{table}

\subsubsection{Case 2: Dissimilar linear[3]catenane}
Following a similar approach, in the second comparative case, we model three dissimilar linear[3]catenane polymers: (44,24,24), (50,18,18), and (58,10,10), all having $R_h \sim 10 \sigma_p$, as shown in Figure \ref{fig:dissimilar}(b). The decay of the RCF of these three dissimilar linear[3]catenane with time are shown in Figure \ref{fig:dissimilar}(d) together with that of linear[3]catenane containing identical constituent rings, i.e., (31,31,31) and the ring polymer (N=64), both considered earlier in this work. 
Similar to the case of dissimilar linear[2]catenane discussed in section \ref{sec:l2c}, as the size of one of the constituent ring of the linear[3]catenane is made bigger, ensuring approximately the same $R_h$ value of the entire linear[3]catenane, the RCF decay curve gradually moves towards that of the ring polymer and eventually nearly coincides with it for the cases of (50,18,18) and (58,10,10). 
\begin{table}
\centering
\caption{Molecular weights of the constituent rings $\boldsymbol{(N_1,N_2,N_3)}$, Hydrodynamic radius ($\boldsymbol{\bar{R_h} = \langle 1 / R_h \rangle^{-1}}$), Relative shape anisotropy ($\boldsymbol{\kappa^2}$), and Rotational diffusion coefficient ($\boldsymbol{D_r}$) for linear[3]catenanes having approximately the same hydrodynamic radius ($\boldsymbol{R_h \sim 10 \sigma_p}$).}

\begin{tabular}{cccc}
$\boldsymbol{(N_1,N_2,N_3)}$ & $\boldsymbol{\bar{R_h} / \sigma_p}$ & $\boldsymbol{\kappa^2}$ & $\boldsymbol{D_r \tau }$ \\
$(31,31,31)$ & $10.06 \pm 2.2$ & $0.218 \pm 0.119$ & $1.58 \times 10^{-4}$ \\
$(44,24,24)$ & $10 \pm 1.98$ & $0.221 \pm 0.105$ & $2.04 \times 10^{-4}$ \\
$(50,18,18)$ & $9.95 \pm 2.01$ & $0.24 \pm 0.106$ & $2.71 \times 10^{-4}$ \\
$(58,10,10)$ & $9.92 \pm 2.1$ & $0.271 \pm 0.118$ & $2.73 \times 10^{-4}$ \\
Ring ($N=64$) & $9.97 \pm 2.02$ & $0.274 \pm 0.124$ & $2.86 \times 10^{-4}$ \\
\end{tabular}
\label{tbl:l3c}
\end{table}
Thus, the reorientation dynamics of significantly dissimilar linear[3]catenanes are primarily governed by the corresponding larger constituent ring, since the influence of the other two smaller constituent rings on the conformational constraint affecting the rotational dynamics reduces. Consequently, the reorientation dynamics becomes ring-like. This progressive shift in the reorientation dynamics of the linear[3]catenane into ring-like behavior can also be seen in the corresponding $D_r$ values, as shown in the last column of Table \ref{tbl:l3c} and is represented by the green dashed line in Figure \ref{fig:hyperbola}, where the dissimilar linear[3]catenane polymers are represented by their corresponding values of $\kappa^2$. 

Thus, the analysis of these two comparative cases of dissimilar linear[2]catenane and linear[3]catenane confirms that the presence of complex topological constraints in the form of mechanical bonds influences the reorientation dynamics of the polymer chains significantly by slowing down the rate of reorientation. 

The hydrodynamic radius is defined using the average of solvent-mediated hydrodynamic interactions between all pairs of the constituent monomers over all possible conformations of the polymer chain\cite{kirkwood1948intrinsic}, and hence depends on the relative distances of the constituent monomers from each other. Thus, it includes the effect of various aspects of the polymer chain, such as branching, flexibility, and topological constraints, including the mechanical bonds. The results of the translational diffusion from the present work convey that the effects of various aspects of the polymer chain, including the molecular topological constraints, on its translational diffusion are implicitly taken care of by the hydrodynamic radius. In contrast, the rotational diffusion results from the present work imply that, even though $R_h$ is strongly dependent on the molecular topology\cite{vargas2018communication,robertson2006diffusion}, it is not sufficient to explain the reorientation dynamics of these polymer chains. 

In addition to the hydrodynamic radius, relative shape anisotropy, a dimensionless measure of the average shape of the polymer chain, defined from the traceless deviatoric part of the diagonalized gyration tensor\cite{theodorou1985shape}, is a critical parameter in the reorientation dynamics of the macromolecules, as shown in Pattnayak et al.\cite{pattnayak2024diffusion}. The observed inverse relation between $D_r$ and $\kappa^2$ for the star polymers, approximated as a rectangular hyperbola\cite{pattnayak2024diffusion}, suggests that polymer chains with lower values of $\kappa^2$ have higher values of $D_r$ when $R_h$ is approximately constant. The lower value of $\kappa^2$ represents higher symmetry in the distribution of the constituent monomers with respect to the coordinate axes, which leads the entire polymer molecule to reorient faster. The results from the present work indicate that in the absence of any mechanical bonds, the rectangular hyperbola relation between $D_r$ and $\kappa^2$ at constant $R_h$ is obeyed (linear and ring), while in the presence of mechanical bonds in the polymer chains, there are significant deviations from it, as shown in Figure \ref{fig:hyperbola}. Cyclic[3]catenane has three mechanical bonds, the highest among the macromolecules under consideration in the present work, followed by two mechanical bonds in linear[3]catenane. Each of the linear[2]catenane, trefoil knot, and Borromean ring has only one mechanical bond, though of distinct topology. Thus, for the macromolecules with mechanical bonds, the relation between $D_r$ and $\kappa^2$ at constant $R_h$ is approximated as a power law, $D_r \sim (\kappa^2)^{b}$, where $b$ is found to be $-4/3$, shown by red dotted lines in Figure \ref{fig:hyperbola}. This exponent represents the slowing down effect on the reorientation dynamics of macromolecules due to mechanical bonds. The presence of mechanical bonds in the polymer chain puts additional constraints on the movement of the constituent monomers, which hinders the reorientation rate of the polymer chain. Further, when the effect of these mechanical bonds is gradually reduced in the two cases of dissimilar linear[2]catenane and linear[3]catenane at approximately constant $R_h$, the values of their respective rotational diffusion coefficients gently move towards that of the ring polymer, as shown in Figure \ref{fig:hyperbola}. Thus, in the reorientation dynamics of mechanically interlocked macromolecules, there is an explicit effect of the mechanical bonds on the rate of reorientation of the polymer chain in addition to the effect of $\kappa^2$. Hence, there are two geometrical parameters of a topological complex polymer chain that affect its reorientation dynamics simultaneously, apart from its size. The first one is its relative shape anisotropy, the lower value of which fastens its rate of reorientation. The second one is its mechanical bonds, the presence of which slows down its reorientation rate. By considering these two opposing effects, we will compare the reorientation dynamics of the macromolecules considered in the present work with each other, by referring the Figure \ref{fig:hyperbola}. Even though linear[3]catenane, linear[2]catenane, and trefoil knot have less value of $\kappa^2$ than the ring polymer, these all have lower values of $D_r$ than that of the ring polymer due to the presence of the mechanical bonds. Between the Borromean ring and the simple ring, although the former is topologically more constrained due to the presence of a mechanical bond, its $D_r$ value is significantly higher than that of the simple ring polymer due to the Borromean ring having a considerably lower value of shape anisotropy than the simple ring polymer. For the comparative case of the polymers with three constituent rings, i.e., linear[3]catenane, cyclic[3]catenane, and Borromean ring, the decreasing order of their $D_r$ values is Borromean ring $>$ cyclic[3]catenane $>$ linear[3]catenane. This is because the Borromean ring has the lowest value of $\kappa^2$ followed by cyclic[3]catenane and linear[3]catenane. Between the cyclic[3]catenane and the ring polymer, the former has a significantly lower value of $\kappa^2$ (fastens the reorientation rate), and is topologically more constrained due to the presence of three mechanical bonds (slows down the rate of reorientation) than the latter. Thus, a balance between these two opposing effects makes the value of $D_r$ of these two polymers approximately close to each other, as shown in Figure \ref{fig:hyperbola}. These various types of comparative cases among the topologically distinct polymer chains of approximately the same hydrodynamic radius show that the competing interplay between the relative shape anisotropy and the mechanical bonds governs the reorientation dynamics of polymer chains with complex molecular topology when their size is nearly constant.

\section*{Concluding Remarks and Outlook}
In the present work, we have simulated the Brownian diffusion of seven topologically distinct polymer chains having approximately the same hydrodynamic radius using multi-particle collision dynamics. We have shown that the translational diffusion coefficients of these seven polymer chains, having nearly the same hydrodynamic radius, are approximately equal to each other, agreeing very well with the Zimm theory. This implies that the molecular topological constraints do not have any additional effect on translational diffusion dynamics of the polymer chain beyond determining its hydrodynamic radius. Thus, the hydrodynamic radius becomes a critical parameter for studying the translational diffusion dynamics of topologically complex polymer chains. Coming to the reorientation dynamics, we have shown that the rotational diffusion coefficients of the macromolecules with mechanical bonds deviate significantly from the rectangular hyperbola relation between rotational diffusion coefficient and relative shape anisotropy, i.e., $D_r \kappa^2$ = constant, previously observed for star polymers (which are free from any topological constraints) at constant hydrodynamic radius\cite{pattnayak2024diffusion}. This deviation indicates that the complex topological constraints present in the polymer chains in the form of mechanical bonds slow down their rate of reorientation, and hence, approximated as a power law fit, $D_r \propto (\kappa^2)^{b}$ ($b \sim -4/3$). This explicit effect of the mechanical bonds on the reorientation dynamics of the macromolecules is also confirmed by the results from the comparative cases of dissimilar linear[2]catenane and linear[3]catenane at approximately constant hydrodynamic radius. Thus, molecular topology has an explicit effect on the reorientational dynamics of the polymer chain in addition to determining its shape ($\kappa^2$) and size ($R_h$). The comparison study among the topologically distinct macromolecules at approximately constant hydrodynamic radius shows that the competing interplay between the relative shape anisotropy and molecular topological constraints governs the reorientation dynamics of the polymer chains when the size is approximately constant. We believe these fundamental insights into the reorientation dynamics of mechanically interlocked macromolecules obtained from the present work will be useful in various scientific applications, like the rational design of molecular machines\cite{sauvage2008molecular}, in the fingerprinting of single proteins using a nanopore\cite{yusko2017real}, where the rotational diffusion coefficient is one of the key parameters, and in the complex self-assembly kinetics of proteins\cite{newton2015rotational}, where the rotational diffusion affects the dynamical pathway of self-assembly.

\section*{Materials and Methods}

We have used the generic coarse-grained bead-spring model, in which the polymer chain is modeled as a series of identical monomer beads of mass $M$ connected by springs. Good solvent condition is modeled by introducing the excluded volume interactions between the monomers by using the Weeks-Chandler-Andersen (WCA) potential\cite{weeks1971role} as follows,
\begin{equation}
U_{WCA}(r) = 
    \begin{cases}
  4\varepsilon \left[ \left( \frac{\sigma_p}{r} \right)^{12} - \left( \frac{\sigma_p}{r} \right)^6 \right] + \varepsilon & $r $\leq 2^{1/6}\sigma_p$ $ \\
  0 & \text{otherwise}
\end{cases}
\end{equation}
where $r$ is the distance between two monomer beads and $\varepsilon = k_BT$ is the strength of interaction. The adjacent monomer beads are connected by springs with the finitely extensible nonlinear elastic (FENE) potential as follows,
\begin{equation}
    U_{FENE}(r) = 
\begin{cases}
  -\frac{1}{2} k r_0^2 \ln \left[ 1 - \left(\frac{r}{r_0}\right)^2 \right]   & $r $\leq r_0$ $ \\
  \infty & \text{otherwise}
\end{cases}
\end{equation}
where the standard Kremer-Grest parameters\cite{grest1986molecular,kremer1990dynamics} are used for the spring constant ($k = 30 k_BT / \sigma_p^2$) and maximum spring extension ($r_0 = 1.5 \sigma_p$). The polymer chain stiffness is modeled by employing an angular potential\cite{halverson2011molecular,rauscher2018topological} as follows,
\begin{equation} \label{eqn:angle}
    U_{angle}(\theta) = \kappa ( 1 + \cos{ \theta } )
\end{equation}
where $\theta$ is the angle between the bonds connecting three neighboring monomer beads. $\kappa$ controls the stiffness of the polymer chain and equals $1.5k_BT$ following earlier studies\cite{halverson2011molecular,rauscher2018topological}.

We have used a hybrid simulation method, which is a combination of molecular dynamics simulations for the polymer chain (solute) and the multi-particle collision dynamics method\cite{malevanets1999mesoscopic} for modeling solvent-mediated hydrodynamic interactions. In the MPCD technique, the solvent is modeled explicitly as an ensemble of non-interacting point particles of finite mass ($m$). The motion of MPCD particles are governed by alternating ballistic streaming and stochastic collisions. In the streaming step of time interval $\delta t$, the positions ($r_i$) of the MPCD particles are updated as follows,
\begin{equation}
    \boldsymbol{r}_i( t + \delta t) = \boldsymbol{r}_i(t) + \delta t \boldsymbol{v}_i(t)
\end{equation}
where $\boldsymbol{v}_i$ is the particle velocity. During this streaming interval, the positions and velocities of the monomer beads are updated as per Newton's equations of motion, which are integrated numerically using the velocity-Verlet algorithm\cite{frenkel2023understanding,tuckerman1992reversible} with time step $\delta t_{MD}$. During the MPCD collision step, the simulation box is divided into identical cubic cells of size ($a$) equals to diameter of the monomer beads ($a = \sigma_p$), and the velocities ($\boldsymbol{v}_i$) of the particles within the cells are updated stochastically using the momentum-conserving Andersen Thermostat (MPCD-AT)\cite{allahyarov2002mesoscopic}, as follows,
\begin{equation}
    \boldsymbol{v}_i(t + \delta t) = \textbf{v}_{cm}(t) + \boldsymbol{v}_i^{ran} - \Delta \textbf{v}_{cm}^{ran}
\end{equation}
where $\textbf{v}_{cm}$ is the center-of-mass velocity of the collision cell, $\boldsymbol{v}_i^{ran}$ is a random velocity, the three components of which are selected randomly from a Gaussian distribution with variance $k_BT/m$ for the solvent particles and $k_BT/M$ for the monomer beads. $\Delta \textbf{v}_{cm}^{ran}$ is the change in center-of-mass velocity of the collision cell due to addition of $\boldsymbol{v}_i^{ran}$, and subtracting it ensures conservation of linear momentum. This kind of modeling the solvent-monomer interaction by including the monomer beads in the stochastic collision of MPCD is often used in recent studies\cite{hegde2011conformation,jiang2013accurate,nikoubashman2017equilibrium,chen2017effect,chen2018coupling} due to its benefit of avoiding spurious depletion forces\cite{padding2006hydrodynamic}. The collision cells are shifted before every collision step by a vector with the three components chosen randomly from $[-a/2,a/2]$ for ensuring Galilean invariance\cite{ihle2001stochastic}. We have carried out the simulations using the MPCD-AT routines in LAMMPS\cite{plimpton1995fast,LAMMPS}(Chen et al.\cite{chen2018coupling,2017GitHub}).

\begin{figure}[ht]
\centering
\includegraphics[width=0.6\linewidth]{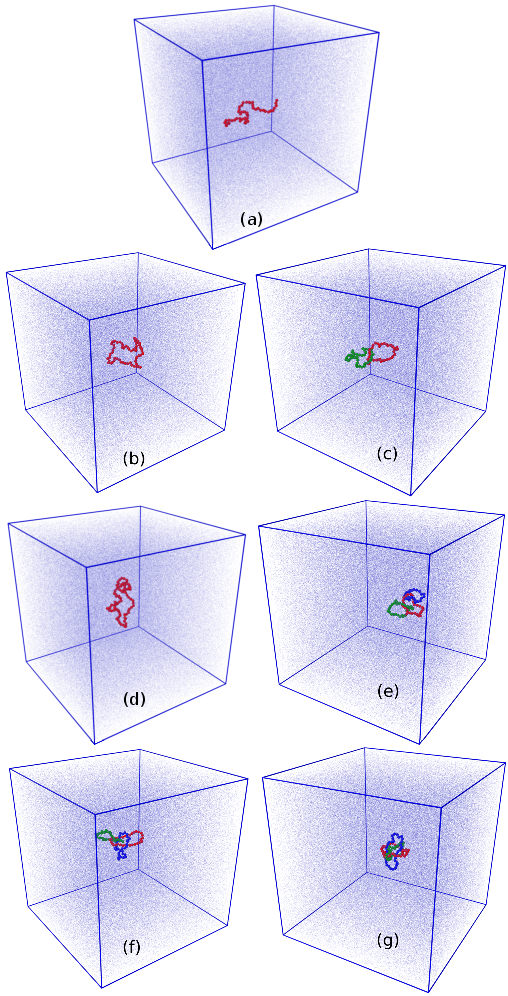}
\caption{MPCD simulation snapshots: (a) linear chain, (b) ring, (c) linear[2]catenane, (d) trefoil knot, (e) linear[3]catenane, (f) cyclic[3]catenane, and (g) Borromean ring. All seven polymer chains have approximately the same hydrodynamic radius ($R_h \sim 10 \sigma_p$). The size of the cubic simulation box is $54 \sigma_p$.}
\label{fig:snapshot}
\end{figure}

The dynamic properties of the MPCD solvent are regulated by the average density of solvent ($\rho_s = 5 m / \sigma_p^3$), and the MPCD time step ($\delta t = 0.09\tau$), where $\tau$ is the intrinsic unit of time equals $\sqrt{m\sigma_p^2/k_BT}$. The mass of a monomer bead ($M$) is selected as 5$m$ for neutral buoyancy. The resulting dynamic viscosity ($\eta$) and the corresponding Schmidt number ($Sc$) of the MPCD solvent are $4 \tau k_BT / \sigma_p^3$ and 12, respectively.
The MD time step ($\delta t_{MD}$) equals $0.002\tau$. The cubic simulation box size is selected to be $54 \sigma_p$ for all constant $R_h$ cases under consideration in order to avoid the finite size effects\cite{hegde2011conformation,chen2017effect}. Periodic boundary conditions are applied in all directions. The equilibration simulation run is performed for $2\times10^7$ MD time steps. The results are time averaged over $5\times10^8$ MD time steps and ensemble-averaged over five system replicas, each with a distinct set of random velocities at the beginning of the simulation and during the stochastic collision, both taken from the Maxwell-Boltzmann distribution. The simulation snapshots are shown in Figure \ref{fig:snapshot}.

\begin{acknowledgement}

G.T. gratefully acknowledges the support received from the National Supercomputing Mission of the Department of Science and Technology (India) for runtime on the PARAM Pravega high-performance computing system housed in the Supercomputing Education and Research Center-Indian Institute of Science, Bengaluru. A.K. and P.K.P. acknowledge partial support received from Anusandhan National Research Foundation (India) grant no. CRG/2022/005381.

\end{acknowledgement}







\bibliography{achemso-demo}

\providecommand{\latin}[1]{#1}
\makeatletter
\providecommand{\doi}
  {\begingroup\let\do\@makeother\dospecials
  \catcode`\{=1 \catcode`\}=2 \doi@aux}
\providecommand{\doi@aux}[1]{\endgroup\texttt{#1}}
\makeatother
\providecommand*\mcitethebibliography{\thebibliography}
\csname @ifundefined\endcsname{endmcitethebibliography}  {\let\endmcitethebibliography\endthebibliography}{}
\begin{mcitethebibliography}{75}
\providecommand*\natexlab[1]{#1}
\providecommand*\mciteSetBstSublistMode[1]{}
\providecommand*\mciteSetBstMaxWidthForm[2]{}
\providecommand*\mciteBstWouldAddEndPuncttrue
  {\def\EndOfBibitem{\unskip.}}
\providecommand*\mciteBstWouldAddEndPunctfalse
  {\let\EndOfBibitem\relax}
\providecommand*\mciteSetBstMidEndSepPunct[3]{}
\providecommand*\mciteSetBstSublistLabelBeginEnd[3]{}
\providecommand*\EndOfBibitem{}
\mciteSetBstSublistMode{f}
\mciteSetBstMaxWidthForm{subitem}{(\alph{mcitesubitemcount})}
\mciteSetBstSublistLabelBeginEnd
  {\mcitemaxwidthsubitemform\space}
  {\relax}
  {\relax}

\bibitem[Brown(1828)]{brown1828xxvii}
Brown,~R. XXVII. A brief account of microscopical observations made in the months of June, July and August 1827, on the particles contained in the pollen of plants; and on the general existence of active molecules in organic and inorganic bodies. \emph{The philosophical magazine} \textbf{1828}, \emph{4}, 161--173\relax
\mciteBstWouldAddEndPuncttrue
\mciteSetBstMidEndSepPunct{\mcitedefaultmidpunct}
{\mcitedefaultendpunct}{\mcitedefaultseppunct}\relax
\EndOfBibitem
\bibitem[Cohen and Moerner(2007)Cohen, and Moerner]{cohen2007principal}
Cohen,~A.~E.; Moerner,~W. Principal-components analysis of shape fluctuations of single DNA molecules. \emph{Proceedings of the National Academy of Sciences} \textbf{2007}, \emph{104}, 12622--12627\relax
\mciteBstWouldAddEndPuncttrue
\mciteSetBstMidEndSepPunct{\mcitedefaultmidpunct}
{\mcitedefaultendpunct}{\mcitedefaultseppunct}\relax
\EndOfBibitem
\bibitem[Gong and van~der Maarel(2014)Gong, and van~der Maarel]{gong2014translational}
Gong,~Z.; van~der Maarel,~J.~R. Translational and reorientational dynamics of entangled DNA. \emph{Macromolecules} \textbf{2014}, \emph{47}, 7230--7237\relax
\mciteBstWouldAddEndPuncttrue
\mciteSetBstMidEndSepPunct{\mcitedefaultmidpunct}
{\mcitedefaultendpunct}{\mcitedefaultseppunct}\relax
\EndOfBibitem
\bibitem[Perkins \latin{et~al.}(1994)Perkins, Smith, and Chu]{perkins1994direct}
Perkins,~T.~T.; Smith,~D.~E.; Chu,~S. Direct observation of tube-like motion of a single polymer chain. \emph{Science} \textbf{1994}, \emph{264}, 819--822\relax
\mciteBstWouldAddEndPuncttrue
\mciteSetBstMidEndSepPunct{\mcitedefaultmidpunct}
{\mcitedefaultendpunct}{\mcitedefaultseppunct}\relax
\EndOfBibitem
\bibitem[Bravo-Anaya \latin{et~al.}(2017)Bravo-Anaya, Mac{\'\i}as, P{\'e}rez-L{\'o}pez, Galliard, Roux, Landazuri, Carvajal~Ramos, Rinaudo, Pignon, and Soltero]{bravo2017supramolecular}
Bravo-Anaya,~L.~M.; Mac{\'\i}as,~E.~R.; P{\'e}rez-L{\'o}pez,~J.~H.; Galliard,~H.; Roux,~D.~C.; Landazuri,~G.; Carvajal~Ramos,~F.; Rinaudo,~M.; Pignon,~F.; Soltero,~J.~A. Supramolecular organization in calf-thymus DNA solutions under flow in dependence with DNA concentration. \emph{Macromolecules} \textbf{2017}, \emph{50}, 8245--8257\relax
\mciteBstWouldAddEndPuncttrue
\mciteSetBstMidEndSepPunct{\mcitedefaultmidpunct}
{\mcitedefaultendpunct}{\mcitedefaultseppunct}\relax
\EndOfBibitem
\bibitem[Smith \latin{et~al.}(1995)Smith, Perkins, and Chu]{smith1995self}
Smith,~D.~E.; Perkins,~T.~T.; Chu,~S. Self-diffusion of an entangled DNA molecule by reptation. \emph{Physical Review Letters} \textbf{1995}, \emph{75}, 4146\relax
\mciteBstWouldAddEndPuncttrue
\mciteSetBstMidEndSepPunct{\mcitedefaultmidpunct}
{\mcitedefaultendpunct}{\mcitedefaultseppunct}\relax
\EndOfBibitem
\bibitem[Kundukad and Van~der Maarel(2010)Kundukad, and Van~der Maarel]{kundukad2010control}
Kundukad,~B.; Van~der Maarel,~J.~R. Control of the flow properties of DNA by topoisomerase II and its targeting inhibitor. \emph{Biophysical journal} \textbf{2010}, \emph{99}, 1906--1915\relax
\mciteBstWouldAddEndPuncttrue
\mciteSetBstMidEndSepPunct{\mcitedefaultmidpunct}
{\mcitedefaultendpunct}{\mcitedefaultseppunct}\relax
\EndOfBibitem
\bibitem[Neuweiler \latin{et~al.}(2009)Neuweiler, Johnson, and Fersht]{neuweiler2009direct}
Neuweiler,~H.; Johnson,~C.~M.; Fersht,~A.~R. Direct observation of ultrafast folding and denatured state dynamics in single protein molecules. \emph{Proceedings of the National Academy of Sciences} \textbf{2009}, \emph{106}, 18569--18574\relax
\mciteBstWouldAddEndPuncttrue
\mciteSetBstMidEndSepPunct{\mcitedefaultmidpunct}
{\mcitedefaultendpunct}{\mcitedefaultseppunct}\relax
\EndOfBibitem
\bibitem[Dobson(2003)]{dobson2003protein}
Dobson,~C.~M. Protein folding and misfolding. \emph{Nature} \textbf{2003}, \emph{426}, 884--890\relax
\mciteBstWouldAddEndPuncttrue
\mciteSetBstMidEndSepPunct{\mcitedefaultmidpunct}
{\mcitedefaultendpunct}{\mcitedefaultseppunct}\relax
\EndOfBibitem
\bibitem[Matthews(1993)]{matthews1993pathways}
Matthews,~C.~R. Pathways of protein folding. \emph{Annual review of biochemistry} \textbf{1993}, \emph{62}, 653--683\relax
\mciteBstWouldAddEndPuncttrue
\mciteSetBstMidEndSepPunct{\mcitedefaultmidpunct}
{\mcitedefaultendpunct}{\mcitedefaultseppunct}\relax
\EndOfBibitem
\bibitem[Rostom \latin{et~al.}(2020)Rostom, White, Yuan, Saito, and Dadmun]{rostom2020polymer}
Rostom,~S.; White,~B.~T.; Yuan,~G.; Saito,~T.; Dadmun,~M.~D. Polymer chain diffusion in all-polymer nanocomposites: Confinement vs chain acceleration. \emph{The Journal of Physical Chemistry C} \textbf{2020}, \emph{124}, 18834--18839\relax
\mciteBstWouldAddEndPuncttrue
\mciteSetBstMidEndSepPunct{\mcitedefaultmidpunct}
{\mcitedefaultendpunct}{\mcitedefaultseppunct}\relax
\EndOfBibitem
\bibitem[Gam \latin{et~al.}(2011)Gam, Meth, Zane, Chi, Wood, Seitz, Winey, Clarke, and Composto]{gam2011macromolecular}
Gam,~S.; Meth,~J.~S.; Zane,~S.~G.; Chi,~C.; Wood,~B.~A.; Seitz,~M.~E.; Winey,~K.~I.; Clarke,~N.; Composto,~R.~J. Macromolecular diffusion in a crowded polymer nanocomposite. \emph{Macromolecules} \textbf{2011}, \emph{44}, 3494--3501\relax
\mciteBstWouldAddEndPuncttrue
\mciteSetBstMidEndSepPunct{\mcitedefaultmidpunct}
{\mcitedefaultendpunct}{\mcitedefaultseppunct}\relax
\EndOfBibitem
\bibitem[Bunning \latin{et~al.}(2000)Bunning, Natarajan, Tondiglia, and Sutherland]{bunning2000holographic}
Bunning,~T.~J.; Natarajan,~L.~V.; Tondiglia,~V.~P.; Sutherland,~R. Holographic polymer-dispersed liquid crystals (H-PDLCs). \emph{Annual Review of Materials Science} \textbf{2000}, \emph{30}, 83--115\relax
\mciteBstWouldAddEndPuncttrue
\mciteSetBstMidEndSepPunct{\mcitedefaultmidpunct}
{\mcitedefaultendpunct}{\mcitedefaultseppunct}\relax
\EndOfBibitem
\bibitem[Coates(1995)]{coates1995polymer}
Coates,~D. Polymer-dispersed liquid crystals. \emph{Journal of Materials Chemistry} \textbf{1995}, \emph{5}, 2063--2072\relax
\mciteBstWouldAddEndPuncttrue
\mciteSetBstMidEndSepPunct{\mcitedefaultmidpunct}
{\mcitedefaultendpunct}{\mcitedefaultseppunct}\relax
\EndOfBibitem
\bibitem[Wolak and Thorne(2013)Wolak, and Thorne]{wolak2013diffusion}
Wolak,~D.~J.; Thorne,~R.~G. Diffusion of macromolecules in the brain: implications for drug delivery. \emph{Molecular pharmaceutics} \textbf{2013}, \emph{10}, 1492--1504\relax
\mciteBstWouldAddEndPuncttrue
\mciteSetBstMidEndSepPunct{\mcitedefaultmidpunct}
{\mcitedefaultendpunct}{\mcitedefaultseppunct}\relax
\EndOfBibitem
\bibitem[Sauvage and Dietrich-Buchecker(2008)Sauvage, and Dietrich-Buchecker]{sauvage2008molecular}
Sauvage,~J.-P.; Dietrich-Buchecker,~C. \emph{Molecular catenanes, rotaxanes and knots: a journey through the world of molecular topology}; John Wiley \& Sons, 2008\relax
\mciteBstWouldAddEndPuncttrue
\mciteSetBstMidEndSepPunct{\mcitedefaultmidpunct}
{\mcitedefaultendpunct}{\mcitedefaultseppunct}\relax
\EndOfBibitem
\bibitem[Sauvage(2017)]{sauvage2017chemical}
Sauvage,~J.-P. From chemical topology to molecular machines (Nobel lecture). \emph{Angewandte Chemie International Edition} \textbf{2017}, \emph{56}, 11080--11093\relax
\mciteBstWouldAddEndPuncttrue
\mciteSetBstMidEndSepPunct{\mcitedefaultmidpunct}
{\mcitedefaultendpunct}{\mcitedefaultseppunct}\relax
\EndOfBibitem
\bibitem[Cot{\'\i} \latin{et~al.}(2009)Cot{\'\i}, Belowich, Liong, Ambrogio, Lau, Khatib, Zink, Khashab, and Stoddart]{coti2009mechanised}
Cot{\'\i},~K.~K.; Belowich,~M.~E.; Liong,~M.; Ambrogio,~M.~W.; Lau,~Y.~A.; Khatib,~H.~A.; Zink,~J.~I.; Khashab,~N.~M.; Stoddart,~J.~F. Mechanised nanoparticles for drug delivery. \emph{Nanoscale} \textbf{2009}, \emph{1}, 16--39\relax
\mciteBstWouldAddEndPuncttrue
\mciteSetBstMidEndSepPunct{\mcitedefaultmidpunct}
{\mcitedefaultendpunct}{\mcitedefaultseppunct}\relax
\EndOfBibitem
\bibitem[Thordarson \latin{et~al.}(2003)Thordarson, Bijsterveld, Rowan, and Nolte]{thordarson2003epoxidation}
Thordarson,~P.; Bijsterveld,~E.~J.; Rowan,~A.~E.; Nolte,~R.~J. Epoxidation of polybutadiene by a topologically linked catalyst. \emph{Nature} \textbf{2003}, \emph{424}, 915--918\relax
\mciteBstWouldAddEndPuncttrue
\mciteSetBstMidEndSepPunct{\mcitedefaultmidpunct}
{\mcitedefaultendpunct}{\mcitedefaultseppunct}\relax
\EndOfBibitem
\bibitem[Davis \latin{et~al.}(2010)Davis, Orlowski, Rahman, and Beer]{davis2010mechanically}
Davis,~J.~J.; Orlowski,~G.~A.; Rahman,~H.; Beer,~P.~D. Mechanically interlocked and switchable molecules at surfaces. \emph{Chemical Communications} \textbf{2010}, \emph{46}, 54--63\relax
\mciteBstWouldAddEndPuncttrue
\mciteSetBstMidEndSepPunct{\mcitedefaultmidpunct}
{\mcitedefaultendpunct}{\mcitedefaultseppunct}\relax
\EndOfBibitem
\bibitem[Kapnistos \latin{et~al.}(2008)Kapnistos, Lang, Vlassopoulos, Pyckhout-Hintzen, Richter, Cho, Chang, and Rubinstein]{kapnistos2008unexpected}
Kapnistos,~M.; Lang,~M.; Vlassopoulos,~D.; Pyckhout-Hintzen,~W.; Richter,~D.; Cho,~D.; Chang,~T.; Rubinstein,~M. Unexpected power-law stress relaxation of entangled ring polymers. \emph{Nature materials} \textbf{2008}, \emph{7}, 997--1002\relax
\mciteBstWouldAddEndPuncttrue
\mciteSetBstMidEndSepPunct{\mcitedefaultmidpunct}
{\mcitedefaultendpunct}{\mcitedefaultseppunct}\relax
\EndOfBibitem
\bibitem[Rauscher \latin{et~al.}(2020)Rauscher, Schweizer, Rowan, and de~Pablo]{rauscher2020dynamics}
Rauscher,~P.~M.; Schweizer,~K.~S.; Rowan,~S.~J.; de~Pablo,~J.~J. Dynamics of poly [n] catenane melts. \emph{The Journal of Chemical Physics} \textbf{2020}, \emph{152}\relax
\mciteBstWouldAddEndPuncttrue
\mciteSetBstMidEndSepPunct{\mcitedefaultmidpunct}
{\mcitedefaultendpunct}{\mcitedefaultseppunct}\relax
\EndOfBibitem
\bibitem[Robertson and Smith(2007)Robertson, and Smith]{robertson2007strong}
Robertson,~R.~M.; Smith,~D.~E. Strong effects of molecular topology on diffusion of entangled DNA molecules. \emph{Proceedings of the National Academy of Sciences} \textbf{2007}, \emph{104}, 4824--4827\relax
\mciteBstWouldAddEndPuncttrue
\mciteSetBstMidEndSepPunct{\mcitedefaultmidpunct}
{\mcitedefaultendpunct}{\mcitedefaultseppunct}\relax
\EndOfBibitem
\bibitem[D{\"u}nweg and Kremer(1991)D{\"u}nweg, and Kremer]{dunweg1991microscopic}
D{\"u}nweg,~B.; Kremer,~K. Microscopic verification of dynamic scaling in dilute polymer solutions: A molecular-dynamics simulation. \emph{Physical review letters} \textbf{1991}, \emph{66}, 2996\relax
\mciteBstWouldAddEndPuncttrue
\mciteSetBstMidEndSepPunct{\mcitedefaultmidpunct}
{\mcitedefaultendpunct}{\mcitedefaultseppunct}\relax
\EndOfBibitem
\bibitem[D{\"u}nweg and Kremer(1993)D{\"u}nweg, and Kremer]{dunweg1993molecular}
D{\"u}nweg,~B.; Kremer,~K. Molecular dynamics simulation of a polymer chain in solution. \emph{The Journal of chemical physics} \textbf{1993}, \emph{99}, 6983--6997\relax
\mciteBstWouldAddEndPuncttrue
\mciteSetBstMidEndSepPunct{\mcitedefaultmidpunct}
{\mcitedefaultendpunct}{\mcitedefaultseppunct}\relax
\EndOfBibitem
\bibitem[D{\"u}nweg \latin{et~al.}(1998)D{\"u}nweg, Grest, and Kremer]{dunweg1998molecular}
D{\"u}nweg,~B.; Grest,~G.~S.; Kremer,~K. Molecular dynamics simulations of polymer systems. \emph{Numerical methods for polymeric systems} \textbf{1998}, 159--195\relax
\mciteBstWouldAddEndPuncttrue
\mciteSetBstMidEndSepPunct{\mcitedefaultmidpunct}
{\mcitedefaultendpunct}{\mcitedefaultseppunct}\relax
\EndOfBibitem
\bibitem[Einstein(1905)]{einstein1905molekularkinetischen}
Einstein,~A. {\"U}ber die von der molekularkinetischen Theorie der W{\"a}rme geforderte Bewegung von in ruhenden Fl{\"u}ssigkeiten suspendierten Teilchen. \emph{Annalen der physik} \textbf{1905}, \emph{4}\relax
\mciteBstWouldAddEndPuncttrue
\mciteSetBstMidEndSepPunct{\mcitedefaultmidpunct}
{\mcitedefaultendpunct}{\mcitedefaultseppunct}\relax
\EndOfBibitem
\bibitem[Wei \latin{et~al.}(2020)Wei, Liu, and Song]{wei2020coarse}
Wei,~J.; Liu,~Y.; Song,~F. Coarse-grained simulation of the translational and rotational diffusion of globular proteins by dissipative particle dynamics. \emph{The Journal of Chemical Physics} \textbf{2020}, \emph{153}\relax
\mciteBstWouldAddEndPuncttrue
\mciteSetBstMidEndSepPunct{\mcitedefaultmidpunct}
{\mcitedefaultendpunct}{\mcitedefaultseppunct}\relax
\EndOfBibitem
\bibitem[Debye(1929)]{debye1929polar}
Debye,~P. Polar molecules. By P. Debye, Ph. D., Pp. 172. New York: Chemical Catalog Co., Inc., 1929. \emph{Journal of the society of chemical industry} \textbf{1929}, \emph{48}, 1036--1037\relax
\mciteBstWouldAddEndPuncttrue
\mciteSetBstMidEndSepPunct{\mcitedefaultmidpunct}
{\mcitedefaultendpunct}{\mcitedefaultseppunct}\relax
\EndOfBibitem
\bibitem[Robertson \latin{et~al.}(2006)Robertson, Laib, and Smith]{robertson2006diffusion}
Robertson,~R.~M.; Laib,~S.; Smith,~D.~E. Diffusion of isolated DNA molecules: Dependence on length and topology. \emph{Proceedings of the National Academy of Sciences} \textbf{2006}, \emph{103}, 7310--7314\relax
\mciteBstWouldAddEndPuncttrue
\mciteSetBstMidEndSepPunct{\mcitedefaultmidpunct}
{\mcitedefaultendpunct}{\mcitedefaultseppunct}\relax
\EndOfBibitem
\bibitem[Kanaeda and Deguchi(2009)Kanaeda, and Deguchi]{kanaeda2009universality}
Kanaeda,~N.; Deguchi,~T. Universality in the diffusion of knots. \emph{Physical Review E-Statistical, Nonlinear, and Soft Matter Physics} \textbf{2009}, \emph{79}, 021806\relax
\mciteBstWouldAddEndPuncttrue
\mciteSetBstMidEndSepPunct{\mcitedefaultmidpunct}
{\mcitedefaultendpunct}{\mcitedefaultseppunct}\relax
\EndOfBibitem
\bibitem[Farimani \latin{et~al.}(2024)Farimani, Ahmadian~Dehaghani, Likos, and Ejtehadi]{farimani2024effects}
Farimani,~R.~A.; Ahmadian~Dehaghani,~Z.; Likos,~C.~N.; Ejtehadi,~M.~R. Effects of linking topology on the shear response of connected ring polymers: Catenanes and bonded rings flow differently. \emph{Physical Review Letters} \textbf{2024}, \emph{132}, 148101\relax
\mciteBstWouldAddEndPuncttrue
\mciteSetBstMidEndSepPunct{\mcitedefaultmidpunct}
{\mcitedefaultendpunct}{\mcitedefaultseppunct}\relax
\EndOfBibitem
\bibitem[Chen and Muthukumar(2021)Chen, and Muthukumar]{chen2021entropic}
Chen,~K.; Muthukumar,~M. Entropic barrier of topologically immobilized DNA in hydrogels. \emph{Proceedings of the National Academy of Sciences} \textbf{2021}, \emph{118}, e2106380118\relax
\mciteBstWouldAddEndPuncttrue
\mciteSetBstMidEndSepPunct{\mcitedefaultmidpunct}
{\mcitedefaultendpunct}{\mcitedefaultseppunct}\relax
\EndOfBibitem
\bibitem[Bonato \latin{et~al.}(2022)Bonato, Marenduzzo, Michieletto, and Orlandini]{bonato2022topological}
Bonato,~A.; Marenduzzo,~D.; Michieletto,~D.; Orlandini,~E. Topological gelation of reconnecting polymers. \emph{Proceedings of the National Academy of Sciences} \textbf{2022}, \emph{119}, e2207728119\relax
\mciteBstWouldAddEndPuncttrue
\mciteSetBstMidEndSepPunct{\mcitedefaultmidpunct}
{\mcitedefaultendpunct}{\mcitedefaultseppunct}\relax
\EndOfBibitem
\bibitem[Li \latin{et~al.}(2024)Li, Zhang, Li, and Puddephatt]{li2024self}
Li,~Z.; Zhang,~J.; Li,~G.; Puddephatt,~R.~J. Self-assembly of the smallest and tightest molecular trefoil knot. \emph{Nature Communications} \textbf{2024}, \emph{15}, 154\relax
\mciteBstWouldAddEndPuncttrue
\mciteSetBstMidEndSepPunct{\mcitedefaultmidpunct}
{\mcitedefaultendpunct}{\mcitedefaultseppunct}\relax
\EndOfBibitem
\bibitem[Li \latin{et~al.}(2024)Li, Wu, Hao, Lei, and Ma]{li2024activity}
Li,~J.-X.; Wu,~S.; Hao,~L.-L.; Lei,~Q.-L.; Ma,~Y.-Q. Activity-driven polymer knotting for macromolecular topology engineering. \emph{Science Advances} \textbf{2024}, \emph{10}, eadr0716\relax
\mciteBstWouldAddEndPuncttrue
\mciteSetBstMidEndSepPunct{\mcitedefaultmidpunct}
{\mcitedefaultendpunct}{\mcitedefaultseppunct}\relax
\EndOfBibitem
\bibitem[Chiarantoni \latin{et~al.}(2025)Chiarantoni, Tagliabue, Mella, and Micheletti]{chiarantoni2025effect}
Chiarantoni,~P.; Tagliabue,~A.; Mella,~M.; Micheletti,~C. Effect of Ring Composition on the Statics and Dynamics of Block Copolyelectrolyte Catenanes. \emph{Macromolecules} \textbf{2025}, \relax
\mciteBstWouldAddEndPunctfalse
\mciteSetBstMidEndSepPunct{\mcitedefaultmidpunct}
{}{\mcitedefaultseppunct}\relax
\EndOfBibitem
\bibitem[Mei \latin{et~al.}(2024)Mei, Grest, Liu, O’Connor, and Schweizer]{mei2024unified}
Mei,~B.; Grest,~G.~S.; Liu,~S.; O’Connor,~T.~C.; Schweizer,~K.~S. Unified understanding of the impact of semiflexibility, concentration, and molecular weight on macromolecular-scale ring diffusion. \emph{Proceedings of the National Academy of Sciences} \textbf{2024}, \emph{121}, e2403964121\relax
\mciteBstWouldAddEndPuncttrue
\mciteSetBstMidEndSepPunct{\mcitedefaultmidpunct}
{\mcitedefaultendpunct}{\mcitedefaultseppunct}\relax
\EndOfBibitem
\bibitem[Yusko \latin{et~al.}(2017)Yusko, Bruhn, Eggenberger, Houghtaling, Rollings, Walsh, Nandivada, Pindrus, Hall, Sept, \latin{et~al.} others]{yusko2017real}
Yusko,~E.~C.; Bruhn,~B.~R.; Eggenberger,~O.~M.; Houghtaling,~J.; Rollings,~R.~C.; Walsh,~N.~C.; Nandivada,~S.; Pindrus,~M.; Hall,~A.~R.; Sept,~D.; others Real-time shape approximation and fingerprinting of single proteins using a nanopore. \emph{Nature nanotechnology} \textbf{2017}, \emph{12}, 360--367\relax
\mciteBstWouldAddEndPuncttrue
\mciteSetBstMidEndSepPunct{\mcitedefaultmidpunct}
{\mcitedefaultendpunct}{\mcitedefaultseppunct}\relax
\EndOfBibitem
\bibitem[Northrup and Erickson(1992)Northrup, and Erickson]{northrup1992kinetics}
Northrup,~S.~H.; Erickson,~H.~P. Kinetics of protein-protein association explained by Brownian dynamics computer simulation. \emph{Proceedings of the National Academy of Sciences} \textbf{1992}, \emph{89}, 3338--3342\relax
\mciteBstWouldAddEndPuncttrue
\mciteSetBstMidEndSepPunct{\mcitedefaultmidpunct}
{\mcitedefaultendpunct}{\mcitedefaultseppunct}\relax
\EndOfBibitem
\bibitem[Newton \latin{et~al.}(2015)Newton, Groenewold, Kegel, and Bolhuis]{newton2015rotational}
Newton,~A.~C.; Groenewold,~J.; Kegel,~W.~K.; Bolhuis,~P.~G. Rotational diffusion affects the dynamical self-assembly pathways of patchy particles. \emph{Proceedings of the National Academy of Sciences} \textbf{2015}, \emph{112}, 15308--15313\relax
\mciteBstWouldAddEndPuncttrue
\mciteSetBstMidEndSepPunct{\mcitedefaultmidpunct}
{\mcitedefaultendpunct}{\mcitedefaultseppunct}\relax
\EndOfBibitem
\bibitem[Zondervan \latin{et~al.}(2007)Zondervan, Kulzer, Berkhout, and Orrit]{zondervan2007local}
Zondervan,~R.; Kulzer,~F.; Berkhout,~G.~C.; Orrit,~M. Local viscosity of supercooled glycerol near T g probed by rotational diffusion of ensembles and single dye molecules. \emph{Proceedings of the National Academy of Sciences} \textbf{2007}, \emph{104}, 12628--12633\relax
\mciteBstWouldAddEndPuncttrue
\mciteSetBstMidEndSepPunct{\mcitedefaultmidpunct}
{\mcitedefaultendpunct}{\mcitedefaultseppunct}\relax
\EndOfBibitem
\bibitem[Li \latin{et~al.}(2009)Li, Wang, and Pielak]{li2009translational}
Li,~C.; Wang,~Y.; Pielak,~G.~J. Translational and rotational diffusion of a small globular protein under crowded conditions. \emph{The Journal of Physical Chemistry B} \textbf{2009}, \emph{113}, 13390--13392\relax
\mciteBstWouldAddEndPuncttrue
\mciteSetBstMidEndSepPunct{\mcitedefaultmidpunct}
{\mcitedefaultendpunct}{\mcitedefaultseppunct}\relax
\EndOfBibitem
\bibitem[Malevanets and Kapral(1999)Malevanets, and Kapral]{malevanets1999mesoscopic}
Malevanets,~A.; Kapral,~R. Mesoscopic model for solvent dynamics. \emph{The Journal of chemical physics} \textbf{1999}, \emph{110}, 8605--8613\relax
\mciteBstWouldAddEndPuncttrue
\mciteSetBstMidEndSepPunct{\mcitedefaultmidpunct}
{\mcitedefaultendpunct}{\mcitedefaultseppunct}\relax
\EndOfBibitem
\bibitem[Allahyarov and Gompper(2002)Allahyarov, and Gompper]{allahyarov2002mesoscopic}
Allahyarov,~E.; Gompper,~G. Mesoscopic solvent simulations: Multiparticle-collision dynamics of three-dimensional flows. \emph{Physical Review E} \textbf{2002}, \emph{66}, 036702\relax
\mciteBstWouldAddEndPuncttrue
\mciteSetBstMidEndSepPunct{\mcitedefaultmidpunct}
{\mcitedefaultendpunct}{\mcitedefaultseppunct}\relax
\EndOfBibitem
\bibitem[Gompper \latin{et~al.}(2009)Gompper, Ihle, Kroll, and Winkler]{gompper2009multi}
Gompper,~G.; Ihle,~T.; Kroll,~D.; Winkler,~R. Multi-particle collision dynamics: A particle-based mesoscale simulation approach to the hydrodynamics of complex fluids. \emph{Advanced computer simulation approaches for soft matter sciences III} \textbf{2009}, 1--87\relax
\mciteBstWouldAddEndPuncttrue
\mciteSetBstMidEndSepPunct{\mcitedefaultmidpunct}
{\mcitedefaultendpunct}{\mcitedefaultseppunct}\relax
\EndOfBibitem
\bibitem[Hegde \latin{et~al.}(2011)Hegde, Chang, Chen, and Khare]{hegde2011conformation}
Hegde,~G.~A.; Chang,~J.-f.; Chen,~Y.-l.; Khare,~R. Conformation and diffusion behavior of ring polymers in solution: A comparison between molecular dynamics, multiparticle collision dynamics, and lattice Boltzmann simulations. \emph{The Journal of chemical physics} \textbf{2011}, \emph{135}, 184901\relax
\mciteBstWouldAddEndPuncttrue
\mciteSetBstMidEndSepPunct{\mcitedefaultmidpunct}
{\mcitedefaultendpunct}{\mcitedefaultseppunct}\relax
\EndOfBibitem
\bibitem[Mussawisade \latin{et~al.}(2005)Mussawisade, Ripoll, Winkler, and Gompper]{mussawisade2005dynamics}
Mussawisade,~K.; Ripoll,~M.; Winkler,~R.; Gompper,~G. Dynamics of polymers in a particle-based mesoscopic solvent. \emph{The Journal of chemical physics} \textbf{2005}, \emph{123}\relax
\mciteBstWouldAddEndPuncttrue
\mciteSetBstMidEndSepPunct{\mcitedefaultmidpunct}
{\mcitedefaultendpunct}{\mcitedefaultseppunct}\relax
\EndOfBibitem
\bibitem[Pattnayak \latin{et~al.}(2024)Pattnayak, Kumar, and Tomar]{pattnayak2024diffusion}
Pattnayak,~P.~K.; Kumar,~A.; Tomar,~G. Diffusion dynamics of star-shaped macromolecules in dilute solutions. \emph{Macromolecules} \textbf{2024}, \emph{57}, 6657--6665\relax
\mciteBstWouldAddEndPuncttrue
\mciteSetBstMidEndSepPunct{\mcitedefaultmidpunct}
{\mcitedefaultendpunct}{\mcitedefaultseppunct}\relax
\EndOfBibitem
\bibitem[Theodorou and Suter(1985)Theodorou, and Suter]{theodorou1985shape}
Theodorou,~D.~N.; Suter,~U.~W. Shape of unperturbed linear polymers: polypropylene. \emph{Macromolecules} \textbf{1985}, \emph{18}, 1206--1214\relax
\mciteBstWouldAddEndPuncttrue
\mciteSetBstMidEndSepPunct{\mcitedefaultmidpunct}
{\mcitedefaultendpunct}{\mcitedefaultseppunct}\relax
\EndOfBibitem
\bibitem[Wong \latin{et~al.}(2009)Wong, Case, and Szabo]{wong2009influence}
Wong,~V.; Case,~D.~A.; Szabo,~A. Influence of the coupling of interdomain and overall motions on NMR relaxation. \emph{Proceedings of the National Academy of Sciences} \textbf{2009}, \emph{106}, 11016--11021\relax
\mciteBstWouldAddEndPuncttrue
\mciteSetBstMidEndSepPunct{\mcitedefaultmidpunct}
{\mcitedefaultendpunct}{\mcitedefaultseppunct}\relax
\EndOfBibitem
\bibitem[Perrin(1934)]{perrin1934movement}
Perrin,~F. Brownian motion of an ellipsoid-I. Dielectric dispersion for ellipsoidal molecules. \emph{Journal of Physics and Radium} \textbf{1934}, \emph{5}, 497--511\relax
\mciteBstWouldAddEndPuncttrue
\mciteSetBstMidEndSepPunct{\mcitedefaultmidpunct}
{\mcitedefaultendpunct}{\mcitedefaultseppunct}\relax
\EndOfBibitem
\bibitem[Perrin(1936)]{perrin1936movement}
Perrin,~F. Brownian motion of an ellipsoid (II). Free rotation and depolarization of fluorescence. Translation and diffusion of ellipsoidal molecules. \emph{Journal of Physics and Radium} \textbf{1936}, \emph{7}, 1--11\relax
\mciteBstWouldAddEndPuncttrue
\mciteSetBstMidEndSepPunct{\mcitedefaultmidpunct}
{\mcitedefaultendpunct}{\mcitedefaultseppunct}\relax
\EndOfBibitem
\bibitem[Ryabov \latin{et~al.}(2006)Ryabov, Geraghty, Varshney, and Fushman]{ryabov2006efficient}
Ryabov,~Y.~E.; Geraghty,~C.; Varshney,~A.; Fushman,~D. An efficient computational method for predicting rotational diffusion tensors of globular proteins using an ellipsoid representation. \emph{Journal of the American Chemical Society} \textbf{2006}, \emph{128}, 15432--15444\relax
\mciteBstWouldAddEndPuncttrue
\mciteSetBstMidEndSepPunct{\mcitedefaultmidpunct}
{\mcitedefaultendpunct}{\mcitedefaultseppunct}\relax
\EndOfBibitem
\bibitem[{\v{S}}olc(1971)]{vsolc1971shape}
{\v{S}}olc,~K. Shape of a Random-Flight Chain. \emph{The Journal of Chemical Physics} \textbf{1971}, \emph{55}, 335--344\relax
\mciteBstWouldAddEndPuncttrue
\mciteSetBstMidEndSepPunct{\mcitedefaultmidpunct}
{\mcitedefaultendpunct}{\mcitedefaultseppunct}\relax
\EndOfBibitem
\bibitem[Khabaz and Khare(2014)Khabaz, and Khare]{khabaz2014effect}
Khabaz,~F.; Khare,~R. Effect of chain architecture on the size, shape, and intrinsic viscosity of chains in polymer solutions: A molecular simulation study. \emph{The Journal of chemical physics} \textbf{2014}, \emph{141}\relax
\mciteBstWouldAddEndPuncttrue
\mciteSetBstMidEndSepPunct{\mcitedefaultmidpunct}
{\mcitedefaultendpunct}{\mcitedefaultseppunct}\relax
\EndOfBibitem
\bibitem[Kirkwood and Riseman(1948)Kirkwood, and Riseman]{kirkwood1948intrinsic}
Kirkwood,~J.~G.; Riseman,~J. The intrinsic viscosities and diffusion constants of flexible macromolecules in solution. \emph{The Journal of Chemical Physics} \textbf{1948}, \emph{16}, 565--573\relax
\mciteBstWouldAddEndPuncttrue
\mciteSetBstMidEndSepPunct{\mcitedefaultmidpunct}
{\mcitedefaultendpunct}{\mcitedefaultseppunct}\relax
\EndOfBibitem
\bibitem[Vargas-Lara \latin{et~al.}(2018)Vargas-Lara, Pazmi{\~n}o~Betancourt, and Douglas]{vargas2018communication}
Vargas-Lara,~F.; Pazmi{\~n}o~Betancourt,~B.~A.; Douglas,~J.~F. Communication: A comparison between the solution properties of knotted ring and star polymers. \emph{The Journal of chemical physics} \textbf{2018}, \emph{149}\relax
\mciteBstWouldAddEndPuncttrue
\mciteSetBstMidEndSepPunct{\mcitedefaultmidpunct}
{\mcitedefaultendpunct}{\mcitedefaultseppunct}\relax
\EndOfBibitem
\bibitem[Weeks \latin{et~al.}(1971)Weeks, Chandler, and Andersen]{weeks1971role}
Weeks,~J.~D.; Chandler,~D.; Andersen,~H.~C. Role of repulsive forces in determining the equilibrium structure of simple liquids. \emph{The Journal of chemical physics} \textbf{1971}, \emph{54}, 5237--5247\relax
\mciteBstWouldAddEndPuncttrue
\mciteSetBstMidEndSepPunct{\mcitedefaultmidpunct}
{\mcitedefaultendpunct}{\mcitedefaultseppunct}\relax
\EndOfBibitem
\bibitem[Grest and Kremer(1986)Grest, and Kremer]{grest1986molecular}
Grest,~G.~S.; Kremer,~K. Molecular dynamics simulation for polymers in the presence of a heat bath. \emph{Physical Review A} \textbf{1986}, \emph{33}, 3628\relax
\mciteBstWouldAddEndPuncttrue
\mciteSetBstMidEndSepPunct{\mcitedefaultmidpunct}
{\mcitedefaultendpunct}{\mcitedefaultseppunct}\relax
\EndOfBibitem
\bibitem[Kremer and Grest(1990)Kremer, and Grest]{kremer1990dynamics}
Kremer,~K.; Grest,~G.~S. Dynamics of entangled linear polymer melts: A molecular-dynamics simulation. \emph{The Journal of Chemical Physics} \textbf{1990}, \emph{92}, 5057--5086\relax
\mciteBstWouldAddEndPuncttrue
\mciteSetBstMidEndSepPunct{\mcitedefaultmidpunct}
{\mcitedefaultendpunct}{\mcitedefaultseppunct}\relax
\EndOfBibitem
\bibitem[Halverson \latin{et~al.}(2011)Halverson, Lee, Grest, Grosberg, and Kremer]{halverson2011molecular}
Halverson,~J.~D.; Lee,~W.~B.; Grest,~G.~S.; Grosberg,~A.~Y.; Kremer,~K. Molecular dynamics simulation study of nonconcatenated ring polymers in a melt. I. Statics. \emph{The Journal of chemical physics} \textbf{2011}, \emph{134}\relax
\mciteBstWouldAddEndPuncttrue
\mciteSetBstMidEndSepPunct{\mcitedefaultmidpunct}
{\mcitedefaultendpunct}{\mcitedefaultseppunct}\relax
\EndOfBibitem
\bibitem[Rauscher \latin{et~al.}(2018)Rauscher, Rowan, and de~Pablo]{rauscher2018topological}
Rauscher,~P.~M.; Rowan,~S.~J.; de~Pablo,~J.~J. Topological effects in isolated poly [n] catenanes: Molecular dynamics simulations and Rouse mode analysis. \emph{ACS Macro Letters} \textbf{2018}, \emph{7}, 938--943\relax
\mciteBstWouldAddEndPuncttrue
\mciteSetBstMidEndSepPunct{\mcitedefaultmidpunct}
{\mcitedefaultendpunct}{\mcitedefaultseppunct}\relax
\EndOfBibitem
\bibitem[Frenkel and Smit(2023)Frenkel, and Smit]{frenkel2023understanding}
Frenkel,~D.; Smit,~B. \emph{Understanding molecular simulation: from algorithms to applications}; Elsevier, 2023\relax
\mciteBstWouldAddEndPuncttrue
\mciteSetBstMidEndSepPunct{\mcitedefaultmidpunct}
{\mcitedefaultendpunct}{\mcitedefaultseppunct}\relax
\EndOfBibitem
\bibitem[Tuckerman \latin{et~al.}(1992)Tuckerman, Berne, and Martyna]{tuckerman1992reversible}
Tuckerman,~M.; Berne,~B.~J.; Martyna,~G.~J. Reversible multiple time scale molecular dynamics. \emph{The Journal of chemical physics} \textbf{1992}, \emph{97}, 1990--2001\relax
\mciteBstWouldAddEndPuncttrue
\mciteSetBstMidEndSepPunct{\mcitedefaultmidpunct}
{\mcitedefaultendpunct}{\mcitedefaultseppunct}\relax
\EndOfBibitem
\bibitem[Jiang \latin{et~al.}(2013)Jiang, Watari, and Larson]{jiang2013accurate}
Jiang,~L.; Watari,~N.; Larson,~R.~G. How accurate are stochastic rotation dynamics simulations of polymer dynamics? \emph{Journal of Rheology} \textbf{2013}, \emph{57}, 1177--1194\relax
\mciteBstWouldAddEndPuncttrue
\mciteSetBstMidEndSepPunct{\mcitedefaultmidpunct}
{\mcitedefaultendpunct}{\mcitedefaultseppunct}\relax
\EndOfBibitem
\bibitem[Nikoubashman and Howard(2017)Nikoubashman, and Howard]{nikoubashman2017equilibrium}
Nikoubashman,~A.; Howard,~M.~P. Equilibrium dynamics and shear rheology of semiflexible polymers in solution. \emph{Macromolecules} \textbf{2017}, \emph{50}, 8279--8289\relax
\mciteBstWouldAddEndPuncttrue
\mciteSetBstMidEndSepPunct{\mcitedefaultmidpunct}
{\mcitedefaultendpunct}{\mcitedefaultseppunct}\relax
\EndOfBibitem
\bibitem[Chen \latin{et~al.}(2017)Chen, Zhao, and Hou]{chen2017effect}
Chen,~A.; Zhao,~N.; Hou,~Z. The effect of hydrodynamic interactions on nanoparticle diffusion in polymer solutions: a multiparticle collision dynamics study. \emph{Soft matter} \textbf{2017}, \emph{13}, 8625--8635\relax
\mciteBstWouldAddEndPuncttrue
\mciteSetBstMidEndSepPunct{\mcitedefaultmidpunct}
{\mcitedefaultendpunct}{\mcitedefaultseppunct}\relax
\EndOfBibitem
\bibitem[Chen \latin{et~al.}(2018)Chen, Poling-Skutvik, Nikoubashman, Howard, Conrad, and Palmer]{chen2018coupling}
Chen,~R.; Poling-Skutvik,~R.; Nikoubashman,~A.; Howard,~M.~P.; Conrad,~J.~C.; Palmer,~J.~C. Coupling of nanoparticle dynamics to polymer center-of-mass motion in semidilute polymer solutions. \emph{Macromolecules} \textbf{2018}, \emph{51}, 1865--1872\relax
\mciteBstWouldAddEndPuncttrue
\mciteSetBstMidEndSepPunct{\mcitedefaultmidpunct}
{\mcitedefaultendpunct}{\mcitedefaultseppunct}\relax
\EndOfBibitem
\bibitem[Padding and Louis(2006)Padding, and Louis]{padding2006hydrodynamic}
Padding,~J.; Louis,~A. Hydrodynamic interactions and Brownian forces in colloidal suspensions: Coarse-graining over time and length scales. \emph{Physical Review E} \textbf{2006}, \emph{74}, 031402\relax
\mciteBstWouldAddEndPuncttrue
\mciteSetBstMidEndSepPunct{\mcitedefaultmidpunct}
{\mcitedefaultendpunct}{\mcitedefaultseppunct}\relax
\EndOfBibitem
\bibitem[Ihle and Kroll(2001)Ihle, and Kroll]{ihle2001stochastic}
Ihle,~T.; Kroll,~D. Stochastic rotation dynamics: A Galilean-invariant mesoscopic model for fluid flow. \emph{Physical Review E} \textbf{2001}, \emph{63}, 020201\relax
\mciteBstWouldAddEndPuncttrue
\mciteSetBstMidEndSepPunct{\mcitedefaultmidpunct}
{\mcitedefaultendpunct}{\mcitedefaultseppunct}\relax
\EndOfBibitem
\bibitem[Plimpton(1995)]{plimpton1995fast}
Plimpton,~S. Fast parallel algorithms for short-range molecular dynamics. \emph{Journal of computational physics} \textbf{1995}, \emph{117}, 1--19\relax
\mciteBstWouldAddEndPuncttrue
\mciteSetBstMidEndSepPunct{\mcitedefaultmidpunct}
{\mcitedefaultendpunct}{\mcitedefaultseppunct}\relax
\EndOfBibitem
\bibitem[Thompson \latin{et~al.}(2022)Thompson, Aktulga, Berger, Bolintineanu, Brown, Crozier, in~'t Veld, Kohlmeyer, Moore, Nguyen, Shan, Stevens, Tranchida, Trott, and Plimpton]{LAMMPS}
Thompson,~A.~P.; Aktulga,~H.~M.; Berger,~R.; Bolintineanu,~D.~S.; Brown,~W.~M.; Crozier,~P.~S.; in~'t Veld,~P.~J.; Kohlmeyer,~A.; Moore,~S.~G.; Nguyen,~T.~D.; Shan,~R.; Stevens,~M.~J.; Tranchida,~J.; Trott,~C.; Plimpton,~S.~J. {LAMMPS} - a flexible simulation tool for particle-based materials modeling at the atomic, meso, and continuum scales. \emph{Comp. Phys. Comm.} \textbf{2022}, \emph{271}, 108171\relax
\mciteBstWouldAddEndPuncttrue
\mciteSetBstMidEndSepPunct{\mcitedefaultmidpunct}
{\mcitedefaultendpunct}{\mcitedefaultseppunct}\relax
\EndOfBibitem
\bibitem[201(2017)]{2017GitHub}
GitHub - palmergroup/mpcd\textunderscore{}polymer\textunderscore{}colloid: Adaptation of {LAMMPS}' stochastic rotation dynamics module to perform multi-particle collision dynamics simulations of polymer and colloid solutions using an {Andersen} thermostat. https://github.com/palmergroup/mpcd\textunderscore{}polymer\textunderscore{}colloid, 2017; [Online; accessed 2023-01-28]\relax
\mciteBstWouldAddEndPuncttrue
\mciteSetBstMidEndSepPunct{\mcitedefaultmidpunct}
{\mcitedefaultendpunct}{\mcitedefaultseppunct}\relax
\EndOfBibitem
\end{mcitethebibliography}

\end{document}